\documentstyle[12pt,epsf,amssymb]{article}
\newcommand{\msbar}{\overline{\rm MS}}
\newcommand{\ri}{\rm RI-MOM}
\newcommand{\mtms}{m_t^{\scriptsize{\overline{\rm
MS}}}(m_t^{\scriptsize{\overline{\rm MS}}})}
\newcommand{\msms}{m_s^{\scriptsize{\overline{\rm
MS}}}(\mu=2\, \mbox{GeV})}
\newcommand{\bea}{\begin{eqnarray}}
\newcommand{\eea}{\end{eqnarray}}
\newcommand{\beq}{\begin{equation}}
\newcommand{\eeq}{\end{equation}}
\newcommand{\gev}{{\rm GeV}}

\begin{document}
\thispagestyle{empty} 
{\par\raggedleft CPHT-RR 057.0600\par}
{\par\raggedleft FTUV-IFIC-00-0806\par}
{\par\raggedleft RM3-TH/00-11\par}
{\par\raggedleft ROMA-1290/00\par}

\vskip 2.25cm

{\par\centering \textbf{\LARGE \bf A Theoretical
 Prediction of the $B_s$-Meson Lifetime Difference  }\\
\par}

{\par\centering \vskip 1.25 cm\par}
{\par\centering {\large D.~Becirevic, D.~Meloni and A.~Retico}\\
\vskip 0.25cm\par}

{\par\centering \textit{Dip.~di Fisica, Univ.~di Roma ``La 
Sapienza''
and}\\
\textit{INFN, Sezione di Roma, P.le A.~Moro 2, I-00185 Roma, 
Italy.}\\
\vskip 0.3cm\par}

{\par\centering {\large V.~Gim\'enez}\\
\vskip 0.25cm\par}

{\par\centering \textit{Dep.~de F\'isica Te\`orica and IFIC, Univ.~de 
Val\`encia,}\\
\textit{Dr.~Moliner 50, E-46100, Burjassot, Val\`encia, 
Spain}\textsf{\textit{.}}\\
\vskip 0.3cm\par}

{\par\centering {\large V.~Lubicz}\\
\vskip 0.25cm\par}

{\par\centering \textit{Dipartimento di Fisica, Universit\`a di Roma Tre
and INFN, Sezione di Roma Tre}\\
\textit{Via della Vasca Navale 84, I-00146 Rome, Italy}.\\
\vskip 0.3cm\par}

{\par\centering {\large G.~Martinelli}\\
\vskip 0.25cm\par}

{\par\centering \textit{Centre de Physique Th\'eorique de l'\'Ecole Polytechnique,}
\centerline{\sl 91128 Palaiseau-Cedex, France}. \\
\vskip 0.5cm\par}

\vskip 1. cm
\hrule
\begin{abstract}
We present the results of a quenched lattice calculation of the 
operator matrix elements  relevant for predicting  the $B_s$ width 
difference.  Our main result
is  $\left(\Delta\Gamma_{B_s}/\Gamma_{B_s}\right)=  (4.7 \pm 1.5 \pm 1.6)
\times 10^{-2}$, obtained 
from the ratio of matrix elements ${\cal R}(m_b)=
\langle \bar B_s^0\vert Q_{S}\vert B_s^0\rangle/
\langle \bar B_s^0\vert Q_{L}\vert B_s^0\rangle=-0.93(3)^{+0.00}_{-0.01}$.
${\cal R}(m_b)$ was evaluated  from the two relevant 
$B$-parameters $B_{S}^{\msbar} (m_b)  = 0.86(2)^{+0.02}_{-0.03}$ and
$B_{B_s}^{\msbar} (m_b) = 0.91(3)^{+0.00}_{-0.06}$, which we computed in
our simulation.
\end{abstract}
\hrule
\vskip 0.2cm
{\par\centering PACS: 13.75Lb,\ 11.15.Ha,\ 12.38.Gc. \par}
\vskip 0.2 cm 
\newpage
\setcounter{page}{1}
\setcounter{footnote}{0}
\setcounter{equation}{0}

\vspace{3.cm}

\section{Introduction}
\label{sec:intro}
In the Standard Model, the width difference 
$\left(\Delta\Gamma_{B_s}/\Gamma_{B_s}\right)$
of $B_s$ mesons is expected to be rather large and within reach for being
measured in the near future. 
Recent experimental studies~\cite{aleph, schneider} 
already provide  an interesting bound on this quantity. 
In particular, in ref.~\cite{schneider} the limit
$\left(\Delta\Gamma_{B_s}/\Gamma_{B_s}\right)
 < 0.31$ $ (95\% C.L.)$ is quoted~\footnote{ For this estimate,  
 the average $B_s$ decay width 
was  assumed to be the same as for $B_d$ mesons.}. 

\par Theoretically, the prediction of $\left( \Delta
\Gamma/\Gamma\right)_{B_s}$ relies on the use of the operator 
product expansion (OPE), where the large scale is provided by the heavy quark
mass~\cite{shifman}. 
All recent developments, including the calculation
of the next-to-leading order (NLO) perturbative QCD 
corrections, have been discussed in great detail 
in refs.~\cite{beneke}--\cite{nierste}. The theoretical estimates are in the 
range $\left( \Delta \Gamma_{B_s}/\Gamma_{B_s}\right)= (5 \div 15) \%$ and 
 crucially depend on the
size of  relevant hadronic matrix elements which must be computed 
non-perturbatively. 
\par In this paper we present a new lattice calculation
of the main contribution to  $(\Delta\Gamma_{B_s}/\Gamma_{B_s})$.
On the basis of our results, and using the expressions  given  below,  we predict
\bea \frac{\Delta\Gamma_{B_s}}{\Gamma_{B_s}} =  (4.7 \pm 1.5 \pm 1.6) \times 
10^{-2}   \, ,
\label{eq:final}\eea
where the last error is obtained by assuming
an  uncertainty of $30\%$ on  the $1/m_b$ corrections.
 
\par 
We now present the relevant formulae which have been used to get the prediction 
in eq.~(\ref{eq:final}).
Up and including  $1/m_b$ corrections,
the theoretical expression for 
 $\left(\Delta\Gamma_{B_s}/\Gamma_{B_s}\right)$
reads~\cite{beneke}  
\bea
\label{eq:formula1}
{ \Delta\Gamma_{B_s}\over\Gamma_{B_s} }= {G_F^2 m_b^2\over 12 \pi m_{B_s}}
 \left|{ V_{cb} V_{cs}}\right|^2 \tau_{B_s} \biggl( G(z) \langle Q_L (m_b)\rangle - 
 G_S(z) \langle Q_S(m_b) \rangle + \delta_{1/m} \sqrt{1-4z} \biggr) \; , 
\eea
where $z=m_c^2/m_b^2$ and 
$G(z)$ and $G_S(z)$ are functions which have been 
computed in perturbation theory at the next-to-leading order
(NLO)~\cite{nierste}. 
$\langle Q_{L,S} \rangle=
\langle \bar B_s^0\vert Q_{L,S}\vert B_s^0\rangle$ 
are the  hadronic matrix elements  of the  renormalized 
operators  relevant at the lowest order
in the heavy quark expansion
\bea 
\label{basis0}
Q_L &= &\bar b^i \gamma^\mu (1-\gamma_5) s^i \ \bar b^j
 \gamma_\mu (1 - \gamma_5) s^j \cr
&& \cr
Q_S &= &\bar b^i (1-\gamma_5) s^i \ \bar b^j (1 - \gamma_5) s^j \, ,
  \eea
where $i$ and $j$ are colour indices.
The last term in eq.~(\ref{eq:formula1}), $\delta_{1/m} \sqrt{1-4z}$, represents the
contribution of $1/m_b$ corrections.  
\par In order to reduce the uncertainties
of the theoretical predictions, it is convenient to consider the ratio 
of the width and mass differences of the $B_s^0$--$\bar B_s^0$ system 
\bea
{ \Delta \Gamma_ {B_s}\over \Delta m_{B_s}} = {4\pi\over 3}
{m_b^2\over m_W^2} \left|{ V_{cb} V_{cs}\over V_{ts} V_{tb}}\right|^2 {1\over
\eta_B(m_b) S_0(x_t)}  \left( G(z) - G_S(z) 
{\langle Q_S \rangle\over \langle Q_L \rangle} + \tilde \delta_{1/m} 
\right) \; , \label{eq:master} \eea
where $\eta_B(m_b)$ has  been 
computed in perturbation theory 
to NLO~\cite{buras} and  $S_0(x_t)$ is the usual Inami-Lim 
function~\cite{inamilim}.  
Note that,  to make contact 
with ref.~\cite{nierste}, in the above formulae  we used the $\msbar$ coefficient $\eta_B(m_b)$ 
instead of the standard $\eta_B$ of ref.~\cite{buras}.
Consequently, the operators $Q_S$ and $Q_L$ are 
renormalized in the $\msbar$ (NDR) scheme.
We see from eq.~(\ref{eq:master}), that $(\Delta \Gamma_ {B_s}/ \Delta m_{B_s})$ only depends on the
ratio of matrix elements,
\bea
{\cal R}(m_b)= {\langle Q_S \rangle\over \langle Q_L \rangle}
= {\langle \bar B_s^0\vert Q_S( m_b)\vert B_s^0\rangle \over
\langle \bar B_s^0\vert Q_L( m_b)\vert B_s^0\rangle}\ ,
\label{calR} \eea
which may, in principle,  be directly determined on the
lattice. The use of  ${\cal R}$ is particularly  convenient because  
dimensionless quantities are not affected by the uncertainty due to the calibration 
of the lattice spacing. One may also  argue that many systematic errors, 
induced by discretization and quenching, cancel in the ratio of two similar 
amplitudes. 

\par Finally, eq.~(\ref{eq:master}) allows to express 
$(\Delta\Gamma_{B_s}/\Gamma_{B_s})$ in
terms of i) perturbative quantities, encoded in the
overall factor and in the functions $G(z)$ and $G_S(z)$, ii) a lattice measured quantity, ${\cal R}$ and 
iii) $\Delta m_{B_s}$ which  will be hopefully precisely measured in 
the near future: 
\bea
{\Delta \Gamma_ {B_s}\over  \Gamma_ {B_s}} \ = \ \left( \tau_{B_s}  
\Delta m_{B_s} \right)^{\rm (exp.)}\ 
{K} \times \left( G(z) - G_S(z) {\cal R}(m_b) + \tilde \delta_{1/m} \right)
\ ,
\label{eq:future} \eea
where
\bea 
K  =
{4\pi\over 3}
{m_b^2\over m_W^2} \left|{ V_{cb} V_{cs}\over V_{ts} V_{tb}}\right|^2 {1\over
\eta_B(m_b) S_0(x_t)}  \ , 
\eea
Waiting  the  measurement of $\Delta m_{B_s}$, for which only a lower bound
presently exists~\cite{expe}, one can use a modified version of eq.~(\ref{eq:future}),
namely
\bea
\label{eq:present} 
{\Delta \Gamma_ {B_s}\over  \Gamma_ {B_s}} \ = \ \left( \tau_{B_s}  
\Delta m_{B_d} \frac{m_{B_s}}{m_{B_d}} \right)^{\rm (exp.)}\ 
\left| V_{ts} \over V_{td} \right|^2 {K} \cdot \biggl( G(z) - G_S(z)
{\cal R} (m_b) + \tilde \delta_{1/m} \biggr)
\ \xi^2 \ ,
\eea
where 
\bea \xi= \frac{ f_{B_s}
\sqrt{\hat B_{B_s}}}{f_{B_d}\sqrt{\hat B_{B_d}}} \ . \eea
In this way,  besides the quantities discussed above, we only use the experimental
$B_d$-meson mass difference, which is known with a tiny error~\cite{schneider}
\bea
 \left( \Delta m_{B_d} \right)^{\rm (exp.)} = 0.484(15)\, ps^{-1} \ ,\eea
and another ratio of hadronic matrix elements, namely $\xi$, which is
rather accurately determined in lattice simulations~\cite{reviews,ourBBar}.
\par For the following discussion, it is useful to write eq.~(\ref{eq:present})
as (note the ${\cal R}(m_b)$ is negative)
\bea
{\Delta \Gamma_ {B_s}\over  \Gamma_ {B_s}} \ =
 \left[ (0.5 \pm 0.1) - (13.8 \pm 2.8) {\cal R}(m_b) + (15.7 \pm 2.8)
\tilde \delta_{1/m}\right] \times  10^{-2} \, ,\label{eq:num} \eea
where the three contributions correspond to $G(z)$, $-G_S(z) {\cal R}(m_b)$
and $\tilde \delta_{1/m}$ in~(\ref{eq:present}), respectively. 
\par The advantage of using eqs.~(\ref{eq:master}), (\ref{eq:future}) and
(\ref{eq:present})  consists also in the fact
that, in order to predict $\left(\Delta\Gamma_{B_s}/\Gamma_{B_s}\right)$, 
we do not need 
$f_{B_s}$, which enters eq.~(\ref{eq:formula1}) when we express the matrix
elements in terms of $B$-parameters.  There is still a considerable 
uncertainty, indeed,  on $f_{B_s}$, which has been evaluated both in the 
quenched ($f_{B_s}=195 \pm 20$~MeV) and unquenched ($f_{B_s}=245\pm
30$~MeV) case, with a sizeable shift between the two central values~\cite{reviews}. 
Since the  ``unquenched" results are still in their infancy, 
however, we think that
the large  quenching effect should not be taken  too seriously yet.

\par In the numerical evaluation of the width difference,
we have used the values and errors  of the parameters 
given in Tab.~\ref{tab:parameters}. 
For the perturbative quantities $G(z)$ and $G_S(z)$,
 the main uncertainty comes from 
dependence on the renormalization scale, which was varied between
$m_b/2$ and $2 m_b$ in ref.~\cite{nierste} (with $m_b=4.8$~GeV).
For this reason, we found it useless to recompute these coefficients
with $m_b=4.6$~GeV (which is very close to the mass used in~\cite{nierste}).
Instead, we took as central value the average 
$G_S(z)=(G_S^{\mu_1=m_b/2}(z)+G_S^{\mu_1=2\ m_b}(z))/2$ 
(see Tab.~1 of that paper), and as error 
$(G_S^{\mu_1=m_b/2}(z)-G_S^{\mu_1=2\ m_b}(z))/2$. 
The same was made  in the case of $G(z)$.
\par   For the hadronic quantities,
we have used the following values
\bea {\cal R}(m_b) = -0.93(3)^{+0.00}_{-0.01} \ ,  \quad\quad  \xi = 1.16 (7) \ , \label{eq:results} \eea
${\cal R}(m_b)$ is the main results of this lattice study, while the result for
$\xi$ has already been reported in our previous paper~\cite{ourBBar}.
Following ref.~\cite{beneke}, for the $1/m_b$ corrections
 we get
\beq \delta_{1/m} = -2.0 \, \mbox{GeV}^4 \; ,\eeq
using $f_{B_s}=204$~MeV from ref.~\cite{ourBBar}.
This value of $\delta_{1/m}$  
corresponds to $\tilde \delta_{1/m} =-0.55$ (obtained 
with $B_{B_s}^{\msbar} (m_b)=0.91$)~\footnote{ In ref.~\cite{ourBBar},
we gave  $B_{B_s}^{\msbar} (m_b) =
0.92(3)^{+0.00}_{-0.06}$. The tiny difference is due to the fact that there,
in the perturbative
evolution, we used a number of flavours
$n_f=4$ instead  of $n_f=0$ as in the present paper.}.
Since in the estimate  of $\tilde \delta_{1/m}$, the operator matrix elements
were computed  in the  vacuum saturation approximation (VSA), and the radiative
corrections were not included,  we allow it to vary by $30\%$, i.e.
 $\tilde \delta_{1/m} =-0.55\pm 0.17$. 
In the numerical evaluation of the factor $K$, we have used the p\^ole
mass $m_b=4.6$~GeV derived from  the $\msbar$ mass in Tab.~\ref{tab:parameters}
 at the NLO. Using these numbers, from eq.~(\ref{eq:present}) we obtain the result
in eq.~(\ref{eq:final}), where
the last error comes from the uncertainty on $\tilde \delta_{1/m}$. 
\par We have an important remark to make. Eq.~(\ref{eq:num}) shows 
explicitly the  cancellation occurring between  the main contribution, 
proportional
to ${\cal R}(m_b)$, and the $1/m_b$ corrections, proportional to 
$\tilde \delta_{1/m}$. Without the latter, we would have found 
a much larger value,
$\left(\Delta\Gamma_{B_s}/\Gamma_{B_s}\right) \simeq 14 \times 10^{-2}$.
$\left(\Delta\Gamma_{B_s}/\Gamma_{B_s}\right)$
would remain  in the $10\%$ range
even with a sizeable, but smaller, value for the $1/m_b$ term,  
$\tilde \delta_{1/m}=-0.30$ for example. This demonstrates that, 
in spite of the 
progress made in the evaluation of the relevant matrix elements ($\langle
Q_S\rangle$ and $\langle Q_L\rangle$), a precise 
determination of the width difference requires a good control 
of the subleading terms in the $1/m_b$ expansion, which is missing to date.
\begin{table}
\vspace*{-2cm}
\begin{center}
\begin{tabular}{|c|c|} 
\hline
{\phantom{\Large{l}}}\raisebox{+.1cm}{\phantom{\Large{j}}}
{\sl Parameter} & {\sl Value and error} \\ \hline
$m_W$ & 80.41 GeV \\
$m_{B_d}$ & 5.279 GeV \\
$m_{B_s}$ & 5.369 GeV \\
$\tau_{B_s}$ &$ 1.460 \pm 0.056 \, ps$ \\
$\vert V_{cb}\vert$ & $0.0395 \pm 0.0017$ \\
$\vert V_{ts}\vert$ & $0.0386 \pm 0.0013$ \\
$\vert V_{cs}\vert  $ & $0.9759 \pm 0.0005$ \\
$\vert V_{td }\vert  $ & $0.0080 \pm 0.0005$ \\
$\alpha_s(M_Z)$ & $0.118 \pm 0.003$ \\
$\Delta m_{B_d}$ & $0.484 \pm 0.015\,  ps^{-1}$
\\ \hline
$m_t$ & $165 \pm 5$ GeV \\ 
$m_b$ & $4.23 \pm 0.07$ GeV \\
$m_c/m_b$ & $0.29 \pm 0.02$ \\
$m_s $ & $ 110 \pm 20$ MeV \\ 
$\eta_B(m_b)$ & $0.85 \pm 0.02 $\\
\hline
$G(z)$ & $0.030\pm 0.007$ \\
$G_S(z)$ & $0.88 \pm 0.14$ \\ \hline 
\end{tabular}
\caption[]{\label{tab:parameters}{\it Average and errors of 
the main parameters.
When the error is negligible it has been omitted.
The heavy-quark masses ($m_t$, $m_b$ and $m_c$) are the $\msbar$ masses renormalized
at their own values, e.g. $m_t=\mtms$.  $m_s=\msms$ is the 
strange quark mass renormalized in $\msbar$ at  the scale $\mu=2$~GeV.
Its central value and error are not important in the numerical evaluation
of $\Delta\Gamma_{B_s}$.}}
\end{center}
\end{table}  
\par One can also use eq.~(\ref{eq:future}), and combine it with~\cite{expe} 
\bea 
\left( \Delta m_{B_s} \right)^{\rm (exp.)} >  14.6 \ ps^{-1} \ ,  
\eea
to obtain a lower bound on $(\Delta \Gamma_ {B_s}/ \Gamma_{B_s})$.
Given the large uncertainties, this bound is  at present rather weak. 
At the  1-$\sigma$ level we get
\bea 
\frac{\Delta \Gamma_{B_s}}{\Gamma_{B_s}} > 2.5 \times 10^{-2} \,\,\,\, (68 \% \mbox{C.L.}) \ .
\eea
Following ref.~\cite{schneider}, from eq.~(\ref{eq:future}) and the limit
$\left(\Delta \Gamma_{B_s}/\Gamma_{B_s}\right) < 0.31$, we 
could also obtain an upper limit
on $ \Delta m_{B_s}$.  In our case this is  not very interesting, 
 however, since we find a very large  upper bound  of  ${\cal O}(200)\ ps^{-1}$.

\par Our prediction in eq.~(\ref{eq:final}) is in good
agreement with ref.~\cite{nierste}.
It is instead about a factor of two smaller than
the result of ref.~\cite{hiroshima2}.  A  detailed comparison
of the two lattice calculations can be found in sec.~\ref{sec:con}.
\par We stress that the theoretical formulae should be evaluated with 
hadronic parameters computed in a coherent way, within the same lattice 
calculation, and  not from different calculations (the ``Arlequin" 
procedure according to ref.~\cite{ciuchini}), 
since  their values and errors are correlated.
All our lattice results were obtained using a non-perturbatively improved 
action~\cite{Luscher}, and with operators renormalized 
on the lattice with the non-perturbative method of ref.~\cite{np},
as implemented in~\cite{DS=2,bibbia}. 
Our new result is ${\cal R}(m_b)$. For completeness, 
we also present some relevant $B$-parameters  which enter the calculation
of the mixing and width difference
\beq
B_{S}^{\msbar} (m_b)  = 0.86(2)^{+0.02}_{-0.03}\;,\quad   
B_{B_s}^{\msbar} (m_b) =
0.91(3)^{+0.00}_{-0.06} \;.
\eeq
  
\par
The remainder of this paper is as follows: in sec.~\ref{sec:strategy} we
discuss the renormalization of the relevant operators and the calculation of their
matrix elements; the extrapolation to the physical points and the evaluation of   
the statistical and systematic errors are presented in sec.~\ref{sec:phys};
sec.~\ref{sec:con} contains a comparison of our results with other calculations
of the same quantities as well as our conclusions.

\section{Calculation of the Matrix Elements}
\label{sec:strategy}
In this section, we discuss the construction of the renormalized operators
which enter the prediction of $\Delta\Gamma_{B_s}$ and describe the extraction
of their matrix elements in our simulation.  

Besides the operators  in eq.~(\ref{basis0}), we also need 
\bea 
\label{basisc}
{\tilde Q}_S &=& \ \bar b^i  (1- \gamma_{5} )  s^j \, 
 \bar b^j   (1- \gamma_{5} ) s^i \,  ,  \nonumber \\
{Q}_P &=& \ \bar b^i  (1- \gamma_{5} ) s^i \, 
\bar b^j  (1 + \gamma_{5} )  s^j \, ,  \\
{\tilde Q}_P&=& \ \bar b^i  (1- \gamma_{5} ) s^j \, 
 \bar b^j (1 +  \gamma_{5} ) s^i \, . \nonumber
\eea 
The  five operators in  eqs.~(\ref{basis0}) and (\ref{basisc}) form 
a complete basis necessary for the
lattice subtractions which will be discussed later on.  In ref.~\cite{bkns}, a 
new method, that allows the calculation of $\Delta F=2$ amplitudes without
subtractions, has been proposed 
and  feasibility studies are underway. If successful, it will  be obviously applied 
also to the calculation of  $\Delta\Gamma_{B_s}$.
 
The matrix elements which contribute to 
$\Delta\Gamma_{B_s}$   are traditionally 
computed in terms of their value in  the  vacuum 
saturation approximation (VSA), by introducing the so called $B$-parameters. The latter 
 encode the mismatch between  full QCD and  VSA  values. 
There is a certain freedom in defining the
$B$-parameters (see for example the discussion in ref.~\cite{giustiwm}). 
For $Q_S$ and $\tilde Q_S$, 
two equivalent definitions  will be used in the following
\bea 
\label{defB}
\langle  \bar B^0_{q} \vert  {Q_L} (\mu) \vert B^0_{q} 
\rangle 
&=& \frac{8}{3}\ m_{B_s}^2 f_{B_s}^2 \ B_{B_s}(\mu)\,, \cr
{\phantom{\large{l}}}\raisebox{.0cm}{\phantom{\large{j}}}
&&\hfill \cr
\langle  \bar B^0_{q} \vert  {Q}_S (\mu) \vert B^0_{q} 
\rangle 
&=& -\frac{5}{3}\  \left( {m_{B_s}\over m_b(\mu) + m_s(\mu)}\right)^2 
m_{B_s}^2 f_{B_s}^2  \   
B_{S}(\mu) \equiv -\frac{5}{3}  m_{B_s}^2 f_{B_s}^2  B_{S}^\prime (\mu) \, , \cr
{\phantom{\large{l}}}\raisebox{.0cm}{\phantom{\large{j}}}
&&\hfill \cr
\langle  \bar B^0_{q} \vert  {\tilde Q}_S (\mu) \vert B_{q} 
\rangle 
&=& \frac{1}{3}\  \left( {m_{B_s}\over m_b(\mu) + m_s(\mu)}\right)^2 
m_{B_s}^2 f_{B_s}^2  \   
\tilde B_{S}(\mu) \equiv \frac{1}{3} m_{B_s}^2 f_{B_s}^2  
\tilde B_{S}^\prime (\mu) \, . 
\eea
The first definition is the traditional one
that requires, for the physical matrix elements, the knowledge of the quark masses; the second
one may present some advantage, because the matrix elements are derived
using physical quantities only ($m_{B_s}$ and $f_{B_s}$).
The label  $(\mu)$ denotes that  operators and quark masses are  renormalized,
in a given renormalization scheme ($\msbar$ in our case) at the scale $\mu$. 
Since the matrix element of the first operator,  essential for   $\bar B_s^0$--$B^0_s$
mixing,  was studied in detail in our previous paper~\cite{ourBBar}, 
here we only consider the two other relevant operators, namely 
 $Q_S$ and $\tilde Q_S$.
\subsection{Matrix elements from correlation functions}
Before presenting our results, let us recall the main steps necessary
to extract the  matrix elements  in numerical  simulations. 
 As usual, one starts from the (Euclidean) three-point correlation 
functions:
\bea
{\cal C}^{(3)}_i (t_{P_1}, t_{P_2}) =  \displaystyle{\sum_{\vec x, \vec y}}   
\ \langle 0\vert P_5 (\vec x, -t_{P_2}) 
{Q}_i (0)  P_5 (\vec y, t_{P_1}) \vert 0\rangle \ ,
\eea
where $Q_i$ denotes any of the operators enumerated in
eqs.~(\ref{basis0}) and (\ref{basisc}). 
When  the  $Q_i$ and the sources $P_5$ (which we 
choose to be $P_5 = i \bar s  \gamma_5 b$) are sufficiently separated in time, 
the lightest pseudoscalar-meson contribution dominates the correlation functions
and the matrix elements can be extracted  
\bea
{\cal C}^{(3)}_i ( t_{P_1}, t_{P_2})  
\stackrel{t_{P_1}, t_{P_2}\gg 0}{\longrightarrow}\, {\sqrt{{\cal{Z}}_P} 
\over 2 M_{P}}e^{- M_P t_{P_1} }\  \langle \bar P \vert 
{Q}_i(a)  \vert P \rangle\  {\sqrt{{\cal{Z}}_P} \over 2 M_{P}} 
e^{- M_P t_{P_2} } ,
\label{eq : meff}
\eea
where $\sqrt{{\cal{Z}}_P}=\vert \langle 0\vert P_5 \vert P\rangle \vert$. We 
take 
both mesons at rest and   label 
the bare operators on the lattice as ${Q}_i(a)$, to distinguish them from 
their continuum counterparts. The  matching to the continuum
renormalized operators  is what we discuss next. 
\subsection{Operator Matching and Renormalization}
In this subsection, we describe the procedure used to get 
the renormalized operator ${Q}_S(\mu)$, relevant in the
calculation of $\Delta\Gamma_{B_s}$, from the lattice bare operators. 
This is achieved through a two steps procedure:
\begin{itemize}
\item[i)] We define the subtracted  operators
 ${Q^\prime}_S$ and $\tilde {Q}^\prime_{S}$, 
 obeying to the continuum Ward identities  (up to corrections of ${\cal O}(a)
$),  as
\bea
\label{mix1}
&&{ Q^\prime}_S  = {Q}_S(a) + \sum_{i=L, P, \tilde P} \Delta_{i}(g_0^2){Q}_i(a)\ ,\cr
&& \cr
&&{\tilde { Q}^\prime}_S  = 
{\tilde Q}_S(a) + \sum_{i=L, P, \tilde P} \widetilde \Delta_{i}(g_0^2){Q}_i(a)\ .
\eea
The  constants $\Delta_{i}(g_0^2)$, which we calculate using the non-perturbative 
method  discussed in  refs.~\cite{DS=2,bibbia}, 
are listed in Tab.~\ref{tab:0}. 
Note that the mixing~(\ref{mix1}) is a 
lattice artifact (as a consequence of the explicit chiral symmetry
breaking in the Wilson action) and the  subtraction  ensures that 
the resulting operators, 
${ Q^\prime}_{S}$ and $\tilde {Q}^\prime_{S}$, have the same  chiral properties as in the 
continuum. 

In principle, the mixing coefficients
$\Delta_{i}(g_0^2)$ are functions of the bare coupling constant $g_0^2$
only.  Therefore, at fixed lattice spacing, they should be   independent
of the scale at which the operators are renormalized.
In practice, due to some systematic effects, they may depend on the
renormalization scale (which corresponds to the virtuality of the external quark
legs). This induces an uncertainty in the determination of the 
physical matrix elements which will be accounted for in the estimate of the
systematic error.

\item[ii)] 
CPS symmetry allows the mixing of  ${Q}_S$ and ${\tilde Q}_S$  under renormalization. 
This is why we must consider both of them, although our main 
target is the matrix element of the renormalized ${Q}_{S}(\mu)$. 
In the  second step,  the operators are renormalized as:
\bea
\left(
\begin{array}{c}
 {Q}_S (\mu) \\
 {\tilde Q}_S (\mu)\\
\end{array}
\right) \ = \ 
\left(
\begin{array}{cc}
 Z_{22}(\mu) &  Z_{23}(\mu) \\
 Z_{32}(\mu) &  Z_{33}(\mu) \\
\end{array}
\right) \  
\left(
\begin{array}{c}
 { Q^\prime}_S \\
 {\tilde {Q}^\prime}_S \\
\end{array}
\right) \; ,
\label{z23}
\eea
where the structure of  mixing ($Z_{23}\neq Z_{32}\neq 0$) 
is the same  as in the continuum. We compute 
the renormalization matrix  non-perturbatively by using 
the method of ref.~\cite{bibbia}, 
in the Landau  RI-MOM renormalization scheme. The results
 for three values of the renormalization scale, 
$\mu = \{$ $1.9$~GeV, $2.7$~GeV, $3.8$~GeV~$\}$, are  given in 
Tab.~\ref{tab:0}.
\end{itemize}
\begin{table} 
\begin{center} 
\begin{tabular}{|c|c|c|c|c|c|c|} 
\hline
{\phantom{\Large{l}}}\raisebox{+.2cm}{\phantom{\Large{j}}}
{ Scale $ \mu$}  & $\Delta_{L}$   & $\Delta_{P}$ & $\Delta_{\tilde P}$ & 
$\widetilde \Delta_{L}$ & $\widetilde \Delta_{P}$ & $\widetilde \Delta_{\tilde P}$ \\   \hline  \hline
{\phantom{\Large{l}}}\raisebox{+.2cm}{\phantom{\Large{j}}}
1.9~GeV &  0.005(1) & 0.219(8)& -0.016(8)& -0.002(0) & -0.094(3) &0.007(3) \\ \hline
{\phantom{\Large{l}}}\raisebox{-.2cm}{\phantom{\Large{j}}}
 2.7~GeV & 0.003(0) & 0.175(5)& -0.014(2)& -0.001(0) & -0.075(2) &0.005(1) \\ \hline
{\phantom{\Large{l}}}\raisebox{-.2cm}{\phantom{\Large{j}}}
 3.8~GeV & 0.002(1) & 0.189(3)& -0.012(2)& -0.001(0) & -0.081(1) &0.003(1) \\ \hline
\end{tabular}\\
\vspace*{1cm}
\begin{tabular}{|c|c|c|c|c|} 
\hline
{\phantom{\Large{l}}}\raisebox{+.2cm}{\phantom{\Large{j}}}
{ Scale $\mu$}  & $Z_{22}(\mu)$   & $Z_{23}(\mu)$ & $Z_{32}(\mu)$ & $Z_{33}(\mu)$\\   \hline 
\hline
{\phantom{\Large{l}}}\raisebox{+.2cm}{\phantom{\Large{j}}}
 1.9~GeV &  0.237(13) & -0.122(16)& 0.313(1)& 1.018(5) \\ \hline
{\phantom{\Large{l}}}\raisebox{-.2cm}{\phantom{\Large{j}}}
 2.7~GeV &  0.282(12) & -0.128(16)& 0.229(0)& 0.883(3)  \\ \hline
{\phantom{\Large{l}}}\raisebox{-.2cm}{\phantom{\Large{j}}}
 3.8~GeV &  0.332(12) & -0.184(16)& 0.203(1)& 0.902(0)   \\ \hline
\end{tabular}
\vspace*{.8cm}
\caption{\label{tab:0}
{\it Numerical results for the $\Delta(g_0^2)$s 
and the matrix $Z(g_0^2,\mu)$.
 They have been evaluated non-perturbatively in the Landau RI-MOM scheme, 
at $\beta=6/g_0^2=6.2$,  at the 
three different scales  $\mu$ given in the table.}}
\end{center}
\end{table}

{\subsection{Extraction of the B-parameters}}

\begin{figure}[h!]
\begin{center}
\begin{tabular}{@{\hspace{-0.9cm}}c c c}
\hfill & \epsfxsize11.8cm\epsffile{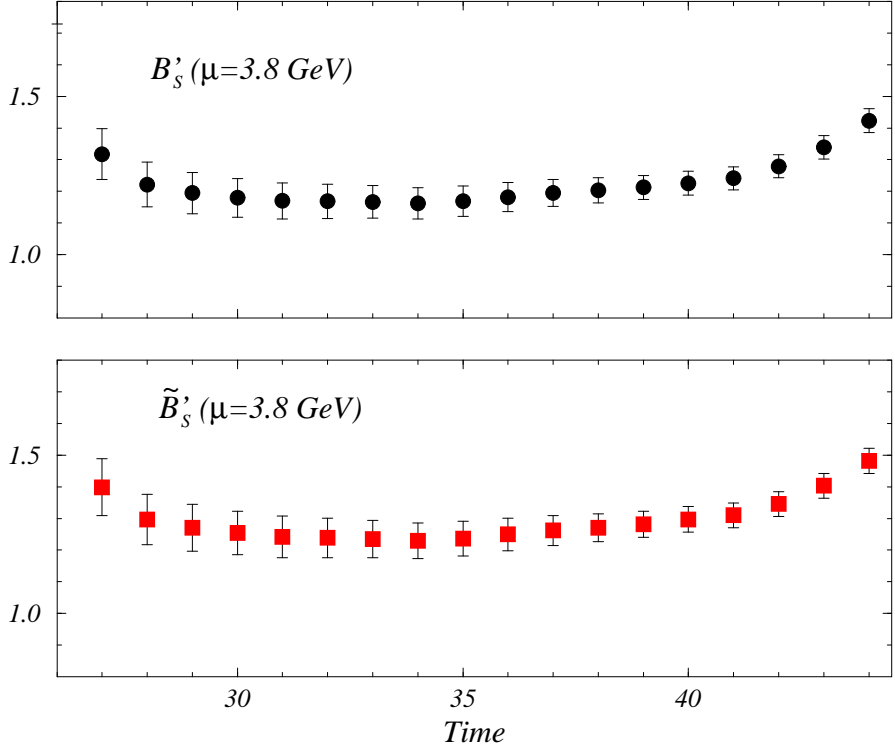} &\hfill  \\
\end{tabular}
\caption{\label{slika1}{\it The ratios defined in eq.~(\ref{method1})
are shown as a function of 
${\rm Time}=-t_{P_2}$ at fixed $t_{P_1}=16$. All quantities are in lattice units. 
The figure refers to  $\kappa_q=0.1349$ and $\kappa_Q = 0.1220$.}}
\end{center}
\end{figure}

Equipped with suitably renormalized operators in the RI-MOM scheme, 
we proceed by removing 
the external meson propagators and sources from the correlation functions.
 This can be done in two ways. From the ratios
\bea
\label{method1}
&& { {\cal C}^{(3)}_S (t_{P_1}, t_{P_2}; \mu) \over \ - {5\over 3}
Z_A^2 {\cal C}_{AP}^{(2)} 
(t_{P_2}) \ {\cal C}_{AP}^{(2)} (t_{P_1})}\ \to \ {\langle \bar P \vert {Q}_S(\mu)  \vert P \rangle \over
- {5\over 3} \vert \langle 0\vert \hat A_0 \vert P\rangle \vert^2} \ \equiv B_S^\prime(\mu) \ ,\cr
 &&\cr 
 &&\cr 
&&{ {\cal C}^{(3)}_{\tilde S} (t_{P_1}, t_{P_2}; \mu) \over \ {1\over 3}
 Z_A^2 {\cal C}_{AP}^{(2)} 
(t_{P_2}) \ {\cal C}_{AP}^{(2)} (t_{P_1})}\ \to \ {\langle \bar P \vert {\tilde Q}_S(\mu)  \vert P \rangle \over
 {1\over 3} \vert \langle 0\vert \hat A_0 \vert P\rangle \vert^2} 
 \ \equiv \tilde B_S^\prime(\mu) 
\eea
we extract the $B^\prime$-parameters ({\sl Method-I}). The quality of 
the resulting plateaus is illustrated in Fig.~\ref{slika1}. 
Since the lattice renormalization constant 
of the axial current ($\hat A_0 =Z_A \bar s \gamma_0\gamma_5 b$) 
is  $\mu$-independent, the anomalous dimension of the parameter 
${B}_S^\prime (\mu)$ is exactly the same as  for
$\langle {Q}_S (\mu)\rangle$.

\begin{figure}[h!]
\begin{center}
\begin{tabular}{@{\hspace{-0.9cm}}c c c}
\hfill & \epsfxsize11.8cm\epsffile{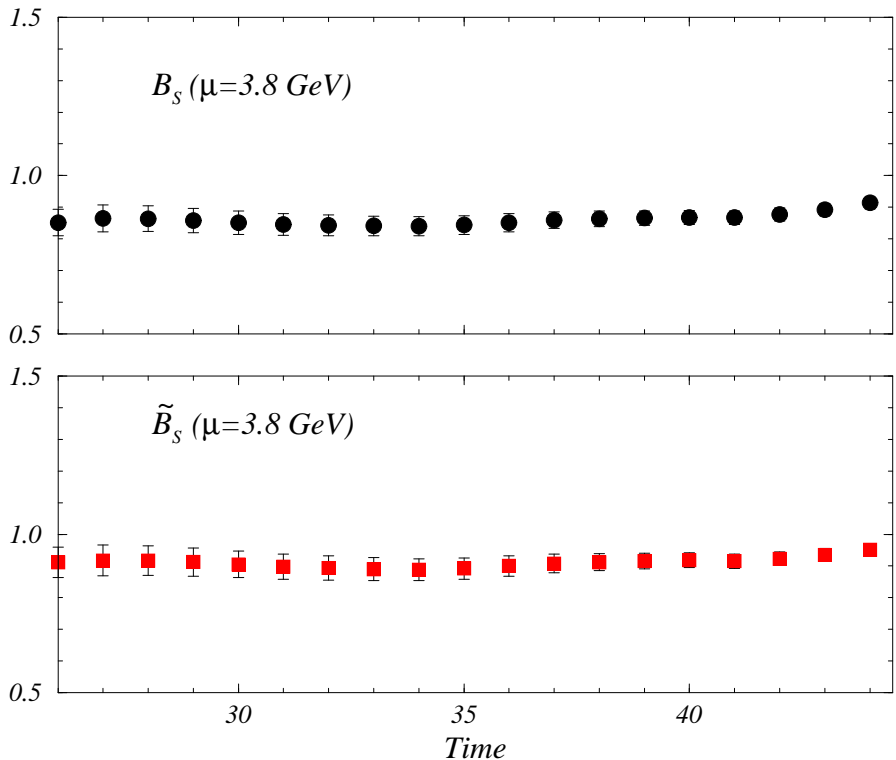} &\hfill  \\
\end{tabular}

\caption{\label{slika2}{\it As in Fig.~{\rm \ref{slika1}}. 
for the standard $B_S$ and $\tilde B_S$,  as specified 
in eq.~(\ref{method2}).}}
\end{center}
\end{figure}

On the other hand, if we combine the correlation functions in the ratios:
\bea
\label{method2}
&& { {\cal C}^{(3)}_S (t_{P_1}, t_{P_2}; \mu) \over \  - {5\over 3}Z^2_P(\mu){\cal C}_{PP}^{(2)} 
(t_{P_2}; \mu) \ {\cal C}_{PP}^{(2)} (t_{P_1}; \mu)}\ \to \  {\langle \bar P \vert {Q}_S(\mu)  \vert P \rangle \over
 - {5\over 3}  \vert \langle 0\vert \hat P_5(\mu) \vert P\rangle \vert^2} \ \equiv B_S (\mu) \ ,\cr
 &&\cr 
 &&\cr 
&&  { {\cal C}^{(3)}_{\tilde S} (t_{P_1}, t_{P_2}; \mu) \over \ {1\over 3}
Z^2_P(\mu) {\cal C}_{PP}^{(2)} 
(t_{P_2}; \mu) \ {\cal C}_{PP}^{(2)} (t_{P_1}; \mu)}\ \to \  {\langle \bar P \vert {\tilde Q}_S(\mu)  \vert P \rangle \over
{1\over 3}  \vert \langle 0\vert \hat  P_5(\mu) \vert P\rangle \vert^2} \ \equiv \tilde B_S (\mu) \ .
\eea
we get the standard $B$-parameters ({\sl Method-II}). 
The plateaus are shown in Fig.~\ref{slika2}.
In the above ratios, we have used the renormalized pseudoscalar density, $\hat P_5(\mu) =i Z_P(\mu) \bar s \gamma_5 b$. 
The value of $Z_P(\mu)$ is also obtained non-perturbatively
in the way described in ref.~\cite{np}.  Numerically (and in the chiral limit), we
have  $Z_P(\mu)=\{ 0.43(2), 0.45(2), 0.50(2)\}$ in increasing order of $\mu$.
\par
The second method has the advantage that the (rather large) $\mu$-dependence of the
operator $\langle {Q}_S (\mu)\rangle$ is almost cancelled by 
the anomalous dimension of the squared renormalized pseudoscalar density.  
On the other hand, the first method  seems more convenient  because 
physical amplitudes can be obtained without introducing the quark masses, which are a further
source of theoretical uncertainty.   We found, however, that  $B^\prime_S$ has 
a very strong dependence 
on the heavy quark mass, which prevents a reliable extrapolation to 
$m_{B_s}$.  Before
discussing the subtleties related to the extrapolation, we present  our results for the
heavy-light meson masses  directly accessible in our simulation.

\subsection{$B$-parameters in the Landau RI-MOM scheme}

\begin{figure}[h!]
\begin{center}
\begin{tabular}{c}
\epsfxsize11.8cm\epsffile{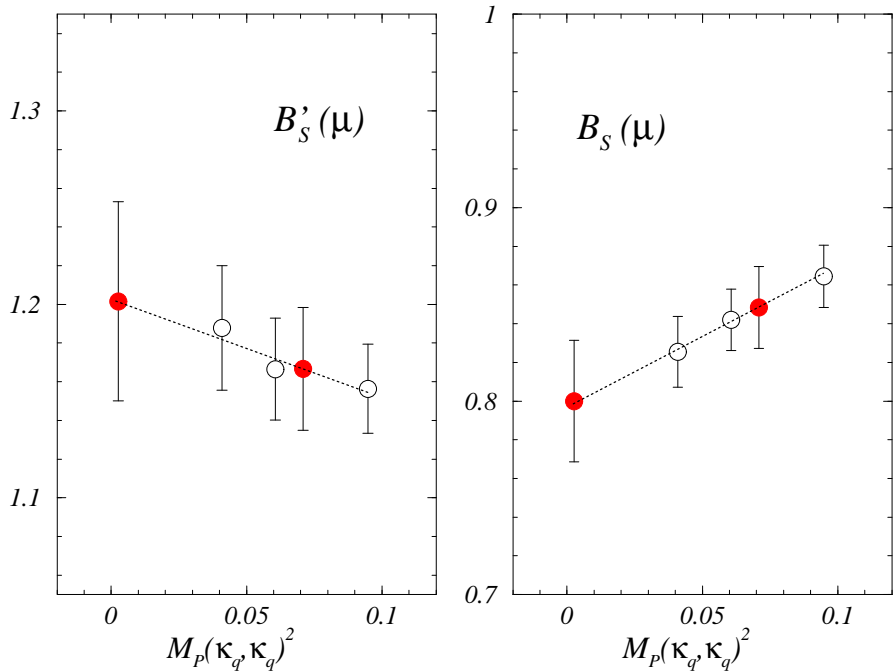}   \\
\end{tabular}

\caption{\label{fig3a}{\it Fit in the light quark mass  of ${B}_S (\mu)$ and
 $ B_S^\prime (\mu)$  
(at $\mu  = 3.8$~GeV) according to eq.~(\ref{lp2}). 
Empty symbols denote   data obtained from our simulation, 
whereas  filled  symbols denote quantities obtained after 
interpolation (extrapolation) to the strange (up-down) light quark mass. 
The heavy-quark mass corresponds  to $\kappa_{Q_2} = 0.1220$.}}
\end{center}
\end{figure}

In this subsection we present the results for  both sets of $B$-parameters.
As in our previous paper~\cite{ourBBar},
our study is based on a sample of $200$ independent quenched gauge field 
configurations, generated at the coupling constant $\beta=6.2$, on the volume $24^3\times
48$. We use three values of the hopping parameter corresponding to the light 
quark mass ($\kappa_q =$ $0.1344$, $0.1349$, $0.1352$), and three values corresponding
to the heavy quarks ($\kappa_Q =$ $0.125$, $0.122$, $0.119$).
The first source is kept fixed at $t_{P_1}=16$, while the second one moves along 
the temporal axis. The 4-fermion operator under study is inserted  at 
the origin ($t=0$). 
After examining the  plateaus of the different  ratios in 
eqs.~(\ref{method1}) and (\ref{method2}),
for every combination of the hopping parameters, we choose to fit in
the time intervals, $t_{B_S} \ \in \ [ 31, 34 ]$ and 
$t_{\tilde B_S} \ \in \ [ 32, 35 ]$.
We  present  results for each value of $\kappa_Q$, 
with  the light quark interpolated to the $s$-quark or  extrapolated to the $d$-quark. 
For a generic $B$-parameter,  this is obtained by fitting our data to the following expression:
\bea
\label{lp2}
{B}(m_Q,m_q) = \alpha_0^Q + \alpha_1^Q   M^2_P(m_q, m_q) \,.
\eea
This  is illustrated in Fig.~\ref{fig3a}, while a detailed list of results 
is presented in Tabs.~\ref{tab:1} and \ref{tab:2}. 
\begin{table}[h!]
\begin{center}
\hspace*{-4mm}
\begin{tabular}{|c|c|c|c|c|c|c|} 
  \hline
{\phantom{\huge{l}}}\raisebox{-.2cm}{\phantom{\Huge{j}}}
{  $\mu$ }&  \multicolumn{2}{c|}{ $1.9$~GeV}  & 
 \multicolumn{2}{c|}{ $2.7$~GeV}  & 
 \multicolumn{2}{c|}{ $3.8$~GeV} \\  \hline 
 {\phantom{\Large{l}}}\raisebox{.2cm}{\phantom{\Large{j}}}
{ Light quark  }&  $q =s$&  $q =d$ &  $q =s$&  $q =d$ &  $q =s$&  $q =d$  \\  \hline \hline
{\phantom{\Large{l}}}\raisebox{+.2cm}{\phantom{\Large{j}}}
$B_S^\prime(\mu;\kappa_Q=0.1250)$ & 0.93(3) & 0.98(5) &
1.15(4) & 1.21(6) & 1.32(4) & 1.38(7)  \\ 
{\phantom{\Large{l}}}\raisebox{-.15cm}{\phantom{\Large{j}}}
$\tilde B_S^\prime(\mu;\kappa_Q=0.1250)$ & {0.99(4)} & { 1.06(6)} & 
{ 1.21(4)} & { 1.29(6)} & { 1.41(5)} & 
{ 1.49(7)}  \\ \hline

{\phantom{\Large{l}}}\raisebox{+.2cm}{\phantom{\Large{j}}}
$B_S^\prime(\mu;\kappa_Q=0.1220)$ & 0.83(3) & 0.85(4) &
1.02(3) & 1.05(5) & 1.17(4) & 1.20(5)  \\ 
{\phantom{\Large{l}}}\raisebox{-.15cm}{\phantom{\Large{j}}}
$\tilde B_S^\prime(\mu;\kappa_Q=0.1220)$ & { 0.87(3)} & { 0.91(5)} & 
{ 1.06(3)} & { 1.10(5)} & { 1.23(4)} & 
{ 1.28(6)}  \\ \hline

{\phantom{\Large{l}}}\raisebox{+.2cm}{\phantom{\Large{j}}}
$B_S^\prime(\mu;\kappa_Q=0.1190)$ & 0.75(2) & 0.77(3) &
0.91(2) & 0.94(4) & 1.05(3) & 1.08(4)  \\ 
{\phantom{\Large{l}}}\raisebox{-.15cm}{\phantom{\Large{j}}}
$\tilde B_S^\prime(\mu;\kappa_Q=0.1190)$ & { 0.77(3)} & { 0.82(4)} & 
{ 0.94(3)} & { 0.99(4)} & { 1.10(3)} & 
{ 1.15(5)}  \\ \hline
\end{tabular}
\vspace*{.8cm}
\caption{\label{tab:1}{\it Results for  $B_S^\prime(\mu)$ and 
$\tilde B_S^\prime(\mu)$ in the Landau RI-MOM scheme. The light 
quark mass is ex\-tra\-po\-la\-ted/in\-ter\-po\-la\-ted to $d/s$ quarks.}}
\end{center}
\end{table}

\begin{table}[h!]
\begin{center}
\hspace*{-4mm}
\begin{tabular}{|c|c|c|c|c|c|c|} 
  \hline
{\phantom{\huge{l}}}\raisebox{-.2cm}{\phantom{\Huge{j}}}
{  $\mu$ }&  \multicolumn{2}{c|}{ $1.9$~GeV}  & 
 \multicolumn{2}{c|}{ $2.7$~GeV}  & 
 \multicolumn{2}{c|}{ $3.8$~GeV} \\  \hline 
 {\phantom{\Large{l}}}\raisebox{.2cm}{\phantom{\Large{j}}}
{ Light quark  }&  $q =s$&  $q =d$ &  $q =s$&  $q =d$ &  $q =s$&  $q =d$  \\  \hline \hline
{\phantom{\Large{l}}}\raisebox{+.2cm}{\phantom{\Large{j}}}
$B_S(\mu;\kappa_Q=0.1250)$ & 0.87(3) & 0.84(4) &
0.85(2) & 0.82(3) & 0.87(2) & 0.79(3)  \\ 
{\phantom{\Large{l}}}\raisebox{-.15cm}{\phantom{\Large{j}}}
$\tilde B_S(\mu;\kappa_Q=0.1250)$ & { 0.92(3)} & { 0.91(4)} & 
{ 0.90(3)} & { 0.88(4)} & { 0.88(3)} & 
{ 0.86(4)}  \\ \hline

{\phantom{\Large{l}}}\raisebox{+.2cm}{\phantom{\Large{j}}}
$B_S(\mu;\kappa_Q=0.1220)$ & 0.90(3) & 0.85(4) &
0.88(2) & 0.83(3) & 0.85(2) & 0.80(3)  \\ 
{\phantom{\Large{l}}}\raisebox{-.15cm}{\phantom{\Large{j}}}
$\tilde B_S(\mu;\kappa_Q=0.1220)$ & { 0.94(3)} & { 0.91(4)} & 
{ 0.92(2)} & { 0.88(4)} & { 0.90(2)} & 
{ 0.86(4)}  \\ \hline

{\phantom{\Large{l}}}\raisebox{+.2cm}{\phantom{\Large{j}}}
$B_S(\mu;\kappa_Q=0.1190)$ & 0.92(2) & 0.91(3) &
0.90(2) & 0.89(3) & 0.87(2) & 0.86(3)  \\ 
{\phantom{\Large{l}}}\raisebox{-.15cm}{\phantom{\Large{j}}}
$\tilde B_S(\mu;\kappa_Q=0.1190)$ &{ 0.95(3)}  & { 0.96(4)} & 
{ 0.93(2)} & { 0.94(3)} & { 0.91(2)} & 
{ 0.91(3)}  \\ \hline
\end{tabular}
\vspace*{.8cm}
\caption{\label{tab:2}{\it As  in Tab.~\ref{tab:1}, but
 for  $B_S(\mu)$ and $\tilde B_S(\mu)$.}}
\end{center}
\end{table}

At this point  we have obtained  the matrix elements parameterized in two ways ( 
$B_S^\prime(\mu)$--$\tilde B_S^\prime(\mu)$  and  $B_S(\mu)$--$\tilde B_S(\mu)$, 
respectively) in the RI-MOM scheme, at three values of the heavy quark masses
 (around the charm quark mass). 
The matrix elements  that we need   refer, instead,
to the $\msbar$ scheme  and 
to the heavy mesons $B_s$. Thus, to get the physical results,
we have to discuss the scheme  dependence, and the extrapolation in
the heavy-meson mass.

\section{The Physical Mixing Amplitudes}
\label{sec:phys}
In this section we discuss the scaling behaviour of the renormalized amplitudes
($B$-parameters), the conversion of the $B$-parameters from the RI-MOM 
to the $\msbar$ scheme and the extrapolation of the results in the heavy-quark
mass. These points are all essential to obtain the final results and to estimate the
systematic errors.
The explicit expressions for the evolution matrices and the 
corrections relating  different schemes have  been derived from 
the results of refs.~\cite{ciuco1,misiakbu}.
\subsection{Scale dependence of the $B$-parameters}
The renormalized operators  obtained
non-perturbatively 
are subject to systematic errors. It is thus important to check whether the 
renormalized matrix elements follow the scaling behaviour predicted by NLO perturbation
theory.  This is also important  because we have finally 
to compute the physical amplitudes by combining  our matrix elements with the
Wilson
coefficients evaluated using perturbation theory in ref.~\cite{beneke} .

The scaling behaviour of the matrix elements is governed by the following equation
(we use the same notation as in ref.~\cite{ciuco1})
\bea
\label{scale2}
 \ \langle \vec {Q} (\mu_2) \rangle \ = \ W^T[\mu_2,\mu_1]^{-1} 
 \langle \vec {Q}  (\mu_1)\rangle \; ,
\eea
where the evolution operator $W[\mu_2,\mu_1]$ can be written as:
\bea
\label{evol}
W[\mu_2,\mu_1] = M(\mu_2) U(\mu_2,\mu_1) M^{-1}(\mu_1) \, .
\eea
 $U(\mu_2,\mu_1)$ is the leading  order  matrix and
the NLO-corrections are encoded in $M(\mu)$. For our purpose, it is convenient 
to rewrite eq.~(\ref{evol}) in the following form
\bea
\label{scale5}
&&W[\mu_2,\mu_1] = w(\mu_2) w^{-1}(\mu_1)\,,  \eea
where 
\bea 
&&  w(\mu)\ =\ M(\mu) \ \alpha_s(\mu)^{-\gamma_0^T/2\beta_0}\ , 
\eea
and $\beta_0 = 11 - 2 n_f/3$. 
The one-loop anomalous dimension is scheme independent and, 
in the basis (\ref{z23}), 
it is given by:
\[
{\gamma_0} =
\left(
\begin{array}{rr}
 -28/3& 4/3 
\\
 16/3 & 32/3 
\\
\end{array}
\right)  \;.
\]
The NLO contribution
\bea
M(\mu)\ =\ \widehat 1\  +\ {J}_{\rm RI-MOM}^{\ n_f}\ {\alpha_s(\mu)\over 4 \pi}\; 
\eea
is given in terms of the  matrix ${J}_{\rm RI-MOM}^{\ n_f}$, \
which we write explicitly for $n_f=0$ 
\bea
&&{J}_{\rm RI-MOM}^{n_f=0}\ = 
 \ \left( \begin{array}{cc}  
{\phantom{\Huge{l}}}\raisebox{-.4cm}{\phantom{\Huge{j}}}
 \frac{\large 170749}{27225} + {44\over 9}\,\log 2&\; -\frac{247372}{27225} + 
 {28 \over 9}\log 2 
\\
{\phantom{\Huge{l}}}\raisebox{-.4cm}{\phantom{\Huge{j}}}
-{{6667}\over {27225}} + {28\over 9}\log 2 &\;  
-{{196424}\over {27225}} + {44\over 9}\log 2
\\
\end{array} \right) \; ,  
\eea
since our lattice results are  obtained in the quenched approximation. 
These formulae are sufficient for the study of the scaling behaviour 
of  $B_S^\prime (\mu)$ and $\tilde B_S^\prime (\mu)$. 
For $B_S (\mu)$ and $\tilde B_S (\mu)$, since they are obtained by  dividing 
the operator matrix
elements by $\langle \hat  P_5(\mu)\rangle  = \langle 0\vert \hat P_5(\mu) \vert P\rangle$, 
we also need the NLO evolution  of the pseudoscalar density
with  the  scale $\mu$,  in the RI-MOM scheme
and with $n_f=0$. This is given by
\beq
\langle \hat P_5^{\rm RI-MOM}(\mu_2) \rangle=
\left( {\alpha_s(\mu_2)\over \alpha_s(\mu_1)}\right)^{-4/11}\ \left( 1 - {489\over 242} {
\alpha_s(\mu_2) - \alpha_s(\mu_1)\over \pi} \right)\langle
\hat  P_5^{\rm RI-MOM}(\mu_1) 
\rangle  \, . 
\eeq

We now use the above formulae   to check whether our lattice results
 scale  as  predicted by perturbation theory.

\begin{figure}[h!]
\begin{center}
\begin{tabular}{c}
\epsfxsize11.2cm\epsffile{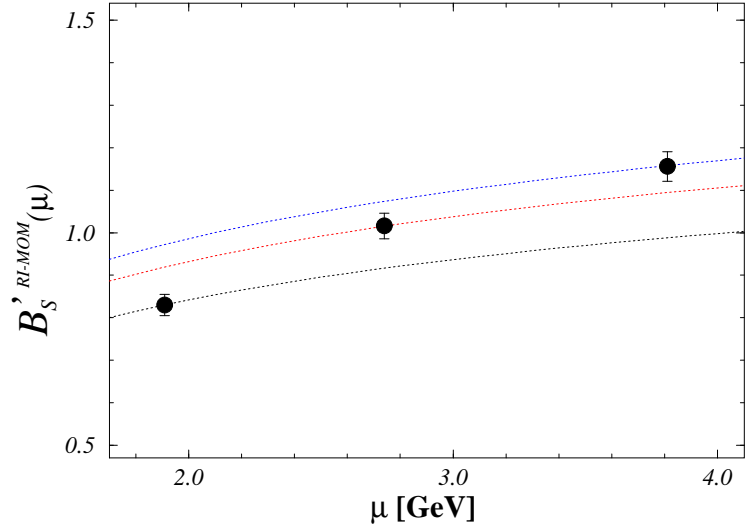} \\
\epsfxsize11.2cm\epsffile{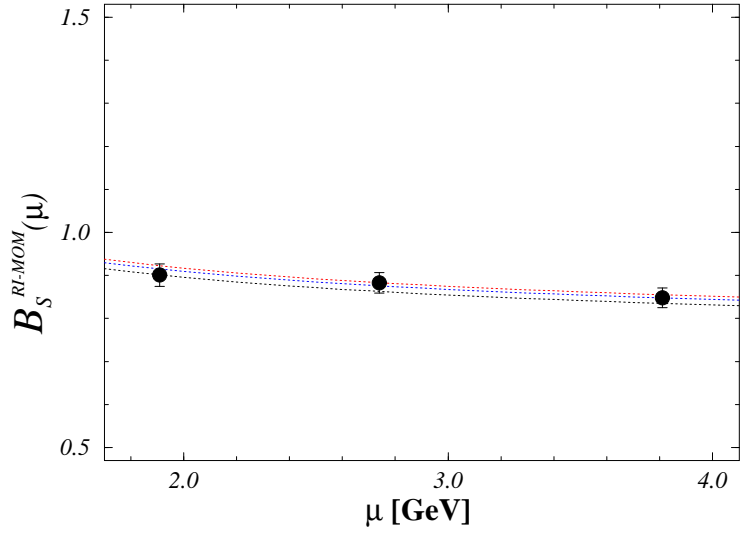} \\
\end{tabular}

\caption{\label{slika3a}{\it NLO evolution of the parameters
  ${B}^\prime_S (\mu)$ and ${B}_S
(\mu)$ in the RI-MOM scheme. The curves  are obtained by starting 
the evolution from the lattice results 
(filled circles) at $\mu  = 1.9$, $2.7$ and $3.8$~{\rm GeV}, respectively.}}
\end{center}
\end{figure}

In Fig.~\ref{slika3a}, we plot the evolution of both $B_S^\prime (\mu)$ and 
$B_S (\mu)$, by normalizing  them at one of the scales 
at which   we have computed the renormalization matrix $Z_{ij}$ in eq.~(\ref{z23}).
From the figure, we see that 
$B_S^\prime(1.9\ {\rm GeV})$ falls below the other  results. On the other hand,
the evolution curves relative to  $B_S^\prime(\mu = 2.7\ {\rm GeV})$ and 
 $B_S^\prime(\mu = 3.8\ {\rm GeV})$ 
are closer to each other. This is to be contrasted to the situation for
$B_S(\mu)$ where the 
scale dependence is not as large as for $B_S^\prime(\mu)$,
and the description of our data by the
perturbative NLO anomalous dimension is more satisfactory, as  also shown in 
Fig.~\ref{slika3a}.
To  convert our result to the $\msbar$ scheme, as central values we choose  
the $B$-parameters obtained with the non-perturbative renormalization
 at $ \mu= 3.8$~GeV. The difference with the other results will be accounted in the
systematic uncertainty. 
\subsection{$B$ parameters in the $\msbar$ scheme at $\mu=m_b$}
In ref.~\cite{beneke},  the  formulae in eqs.~(\ref{eq:formula1}) and (\ref{eq:master})
 were derived in the $\msbar$
scheme. For this reason we have to convert our results from RI-MOM to $\msbar$.
We have chosen to change renormalization scheme before extrapolating in  the 
heavy quark mass. 
The change of scheme is obtained by using the relation 
 \bea
 \langle \vec Q^{\msbar} (\mu) \rangle  
\ =\ \left(\widehat 1\  +\ {r}_{\msbar} \ {\alpha_s(\mu)\over 4 \pi}
\right) \ \langle \vec Q^{\rm RI-MOM} (\mu) 
\rangle \;, \nonumber
\eea
where
\bea
\quad \; {r}_{\msbar} \ =
 \ {1\over 18} \ \left( \begin{array}{cc}  
{\phantom{\Huge{l}}}\raisebox{-.4cm}{\phantom{\Huge{j}}}
  56 + 88\ \log 2&\; -8 + 
 56\ \log 2 
\\
{\phantom{\Huge{l}}}\raisebox{-.4cm}{\phantom{\Huge{j}}}
-143 + 56\ \log 2 &\;  
-115 + 88\ \log 2
\\
\end{array} \right) \ .
\eea 
We note that ${r}_{\msbar}$ is independent of $n_f$.
$ \langle \vec Q^{\msbar} (\mu=3.8 \gev) \rangle$   is then 
evolved  to $\mu = m_b = 4.6$~GeV using 
(\ref{scale2}), with ${J}_{\rm RI-MOM}^{n_f=0}$ replaced 
by
\bea
&& J^{n_f=0}_{\msbar} \ = \ 
 \  
 \left( \begin{array}{cc}  
{\phantom{\Huge{l}}}\raisebox{-.4cm}{\phantom{\Huge{j}}}
 \frac{\large 9561}{3025} 
 &\; -\frac{20723}{18150}
\\
{\phantom{\Huge{l}}}\raisebox{-.4cm}{\phantom{\Huge{j}}}
{{1811}\over {9075}} &\;  
-{{4997}\over {6050}}
\\
\end{array} \right) \  .\eea
The $\msbar$ $B$-parameters at $\mu = m_b^{\rm pole}=  4.6$~GeV
are presented   in Tab.~\ref{tab:2b}.  $m_b^{\rm pole}=  4.6$~GeV corresponds
to  the $\msbar$ mass  given in Tab.~\ref{tab:parameters} and it is very close 
to the value used in~\cite{beneke} to evaluate the Wilson coefficients
at NLO.
\begin{table}[h!]
\begin{center}
\hspace*{-4mm}
\begin{tabular}{|c|c|c|c|c|c|c|} 
  \hline
{\phantom{\huge{l}}}\raisebox{-.2cm}{\phantom{\Huge{j}}}
{  $\kappa_Q$ }&  \multicolumn{2}{c|}{ 0.1250}  & 
 \multicolumn{2}{c|}{0.1220}  & 
 \multicolumn{2}{c|}{ 0.1190 } \\  \hline 
 {\phantom{\Large{l}}}\raisebox{.2cm}{\phantom{\Large{j}}}
{ Light quark  }&  $q =s$&  $q =d$ &  $q =s$&  $q =d$ &  $q =s$&  $q =d$  \\  \hline \hline
 {\phantom{\Large{l}}}\raisebox{.2cm}{\phantom{\Large{j}}}
{ $m_{P}$ [GeV]  }&  1.85(7) &  1.75(8) &  2.11(9) & 2.02(9) &  2.38(10) &  2.26(11)  \\  \hline \hline
{\phantom{\Large{l}}}\raisebox{+.2cm}{\phantom{\Large{j}}}
$B_S^\prime(m_P,m_b)$ & 1.52(5) & 1.60(8) & 1.35(4) & 1.39(7) & 1.21(3) & 1.25(5)  \\ 
{\phantom{\Large{l}}}\raisebox{-.15cm}{\phantom{\Large{j}}}
$\tilde B_S^\prime(m_P,m_b)$ & { 2.18(8)} & { 2.31(12)} & 
{ 1.93(6)} & { 1.99(10)} & { 1.72(5)} & { 1.79(8)}  \\ \hline
{\phantom{\Large{l}}}\raisebox{+.2cm}{\phantom{\Large{j}}}
$B_S(m_P,m_b)$ & 0.73(2) & 0.71(3) &
0.76(2) & 0.72(3) & 0.78(2) & 0.76(3)  \\ 
{\phantom{\Large{l}}}\raisebox{-.15cm}{\phantom{\Large{j}}}
$\tilde B_S(m_P,m_b)$ & { 1.05(3)} & { 1.03(5)} &  { 1.08(3)} & { 1.03(5)} & 
{ 1.10(3)} & { 1.09(5)}  \\ \hline
\end{tabular}
\vspace*{.8cm}
\caption{\label{tab:2b}{\it $B$-parameters in the $\msbar$ scheme
 at $\mu= 4.6$~GeV.}}
\end{center}
\end{table}

\subsection{Extrapolation to the $B_s$ meson}

\begin{figure}[h!]
\begin{center}
\begin{tabular}{c c c}
 &\epsfxsize10.2cm\epsffile{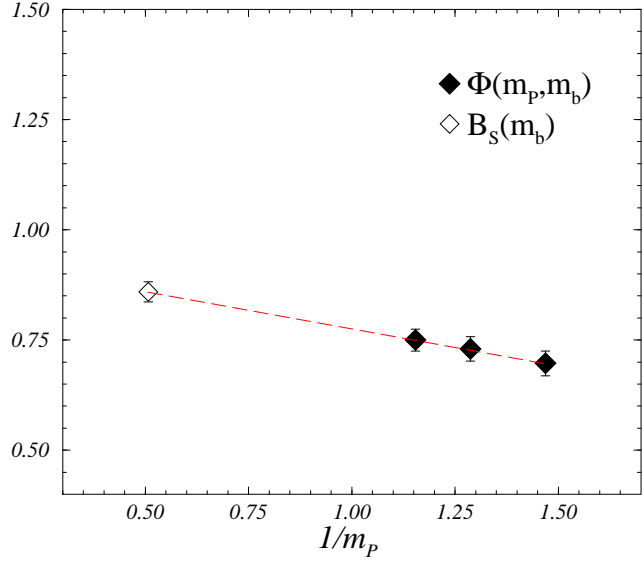} & \\
  &\epsfxsize10.2cm\epsffile{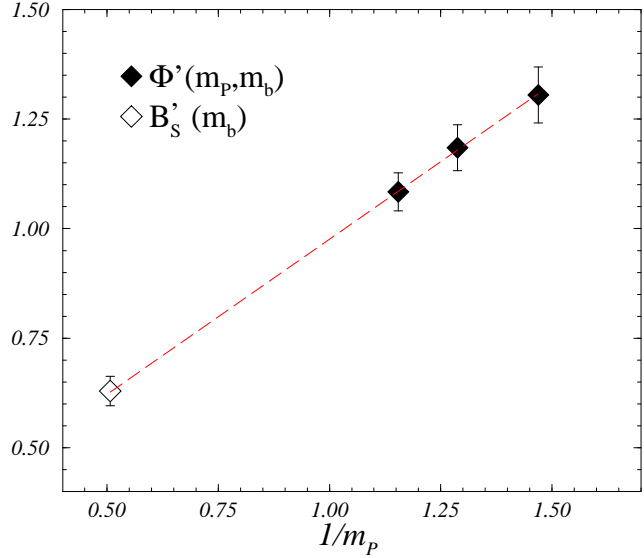}  &\\
\end{tabular}

\caption{\label{fig:extrap}{\it Extrapolation 
of $\Phi(m_P, m_b)$ and $\Phi^\prime(m_P, m_b)$ in $1/m_P$ to 
the $B_s$-meson mass.  $m_P$ is given in lattice units.}}
\end{center}
\end{figure}

From  Tabs.~\ref{tab:1}, \ref{tab:2} and \ref{tab:2b}, we observe that the
dependence of $B_S^\prime$
on both the renormalization scale and the heavy-quark (meson)
mass is much more pronounced than in the case of the  parameter $B_S$
(see also Figs.~\ref{slika3a} and \ref{fig:extrap}). 
In particular, because of the large mass dependence,
it is more difficult to  extrapolate  $B_S^\prime$ to the
physical point.
This is related to the fact that  in eqs.~(\ref{defB}) the ratio 
$(m_{B_s}/m_b)^2$,  distinguishing $B_S$ from $B^\prime_S$,
 still varies
very rapidly in the mass range (around the charm) 
covered in our simulation. A similar problem is found if one tries to extract 
 $m_b$ from the heavy-meson spectrum, computed 
with fully propagating  quarks, by extrapolating in the heavy-quark mass.
For this reason, so far, $m_b$ has been computed on the lattice 
only by using  the  Heavy Quark Effective Theory
(HQET)~\cite{hqet} or Non-Relativistic QCD (NRQCD)~\cite{nrqcd}.
Thus, although in principle we would prefer $B^\prime_S$ because it allows the evaluation
of the matrix element without using the quark mass, in practice our best results
are those obtained from $B_S$.  
For the same reasons, we were unable to extract  
${\cal R}(m_b)= \langle Q_S\rangle /\langle Q_L\rangle$
directly from the ratio of the matrix elements.
The strongest dependence of $B^\prime_S$ on the quark mass  finds its
explanation in the framework of the HQET, by studying the leading and
subleading contributions in the $1/m_b$ expansion
to the $B$ parameters~\cite{reyes}. 
 Our estimates of $\left(\Delta\Gamma_{B_s}/\Gamma_{B_s}\right)$
are then  obtained with ${\cal R}(m_b)$  computed 
by  using $B_S$, $B_{B_s}$
 and $m_b$ from Tab.~\ref{tab:parameters}, as shown below.
For completeness we also present the results  for   $B^\prime_S$.
Hopefully, when smaller values of the lattice spacing, and hence larger values
of the heavy quark mass, will be accessible, 
an accurate determination of the physical matrix element will be provided
by $B^\prime_S$. 

\par In order to make the extrapolation in the  heavy-quark mass from the region 
where we have data to  $m_b$, we rely on 
the scaling laws of the  HQET.  
The  $B$-parameters in Tab.~\ref{tab:2b} have been obtained
using the anomalous dimension matrix of the massless theory. This   is 
the appropriate  procedure since, on our data, $\mu$ is larger than the heavy quark mass,
$m_Q$, used in our simulation. In order to use the HQET scaling laws, 
we have to evolve
first to a scale smaller than the quark mass, at $m_Q$ fixed, and then to
study the scaling at fixed $\mu$ as a function of the heavy quark mass.
In practice, this is achieved, at LO in the anomalous dimensions,
by introducing the following quantities:
\bea
\label{eq:scaling}
&&\vec \Phi (m_{P_s}, m_b) = \left( {\alpha_s(m_{P_s})\over 
\alpha_s(m_{B_s})}\right)^{\gamma/2\beta_0} \vec
B(m_{P_s},m_b)\cr
&& \cr
&&\vec \Phi^\prime (m_{P_s}, m_b) = \left( {\alpha_s(m_{P_s})\over 
\alpha_s(m_{B_s})}\right)^{\gamma^\prime/2\beta_0} \vec
B^\prime(m_{P_s},m_b)
\eea
where we have introduced the matrices $\gamma = \gamma_0 - \widetilde \gamma_0 +
 2 (\widetilde \gamma_P - \gamma_P) $, 
and $\gamma^\prime = \gamma_0 - \widetilde \gamma_0 + 2 \widetilde 
\gamma_A$. 
$\vec B = \left( B_S, \tilde B_S \right)^T$ and similarly for $\vec B^\prime$.
$\widetilde \gamma_0$ is the HQET anomalous dimension matrix
which at leading order is given by
\bea
&&\widetilde \gamma_0 \ = \ -{8\over 3}  
 \  
 \left( \begin{array}{cc}  
2 &\; 1
\\
1 &\; 2
\\
\end{array} \right) \  ,
\eea
whereas  $\widetilde \gamma_P= \widetilde \gamma_A=
-4$ and  the LO anomalous dimension of the pseudoscalar density in full QCD is 
given by  $\gamma_P=-8$.

The quantity $\vec \Phi (m_{P_s}, m_b)$ scales with the heavy quark (heavy meson) as:
\bea
\vec \Phi (m_{P_s}, m_b)\ = \ \alpha + {\beta \over m_{P_s}} + \dots
\eea
\noindent
The extrapolations of  $B_{S}^{\msbar} (m_b)$ and 
$B_{S}^{\prime \msbar} (m_b)$ to  $m_{B_s}$ are shown in Fig.~\ref{fig:extrap}.
The results are:
\bea
&&B_{S}^{\msbar} (m_b)  = 0.86(2)\;,\quad  \tilde B_{S}^{\msbar} (m_b) = 1.25(3) \cr
&& \cr
&&B_{S}^{\prime \ \msbar} (m_b) = 0.63(3)\;,\quad 
\tilde B_{S}^{\prime \ \msbar} (m_b) = 0.91(5) \;.
\eea
As for the systematic uncertainties, we account for the following two:
\begin{itemize}  
\item if, instead of converting to the $\msbar$ scheme before extrapolating to $m_{B_s}$, 
we extrapolate our  RI-MOM results (obtained at $\mu = 3.8$~GeV) in the heavy quark by using 
the same scaling laws of eqs.~(\ref{eq:scaling}), then evolve to $\mu=m_b$, and convert to the
$\msbar$ scheme, the results are~\footnote{If the last conversion from the $\ri$ to the $\msbar$ scheme is made by using $n_f=4$ the results
remain remarkably stable (they decrease by about 1\% ).}:
\bea
&&B_{S}^{\msbar} (m_b)  = 0.88(2)\;,\quad  \tilde B_{S}^{\msbar} (m_b) = 1.27(3) \cr
&& \cr
&&B_{S}^{\prime \ \msbar} (m_b) = 0.62(3)\;,\quad 
\tilde B_{S}^{\prime \ \msbar} (m_b) = 0.90(5) \;.
\eea
\item if we start from  a $\mu$ lower than $3.8$~GeV, for example  $\mu=1.9$~GeV,
then  $B_S(m_b)$ and 
$\tilde B_S(m_b)$ change by $- 3$\%, while $B^\prime_S(m_b)$ and 
$\tilde B^\prime_S(m_b)$ drop by $- 6$\%.
\end{itemize}
After combining these uncertainties, we arrive at our final results:
\bea
B_{S}^{\msbar} (m_b)  = 0.86(2)^{+0.02}_{-0.03}\;,&& \tilde 
B_{S}^{\msbar} (m_b)  =
1.25(3)^{+0.02}_{-0.05}\;, \cr
&&\cr 
B_{S}^{\prime \ \msbar} (m_b)  = 0.63(3)^{+0.00}_{-0.04}\;,&& \tilde
 B_{S}^{\prime \ \msbar} (m_b)  =
0.91(5)^{+0.00}_{-0.06}\; . 
\label{bis} \eea

\par
From these numbers, we observe
 that there is a discrepancy
 between  the value of $B_{S}^{\msbar} (m_b)$ and $B_{S}^{\prime \
  \msbar} (m_b)$.
Using  $B_{S}^{\prime\ \msbar} (m_b) \simeq (m_{B_s}/m_b)^2
B_{S}^{\msbar} (m_b)$,  with $(m_{B_s}/m_b)^2 = 1.6$ from
Tab.~\ref{tab:parameters}, we get $B_{S}^{\prime\ \msbar} (m_b)\sim 1.37$, which is about
twice the value of $B_{S}^{\prime \ \msbar} (m_b)$ in (\ref{bis}).

\par In order to obtain the physical results
we have used  $B_{S}^{\msbar} (m_b)$,
for which the extrapolation to the $B_s$ meson is much smoother.
In this case, we are obliged to use the
quark mass to derive the matrix element 
needed to obtain  ${\cal R}(m_b)$,  which is the  relevant 
quantity for $\left(\Delta\Gamma_{B_s}/\Gamma_{B_s}\right)$, as explained in the introduction. 
This is not a major concern, however, since $m_b$
is known with a very tiny error (see Tab.~\ref{tab:parameters}).

\par The best way to minimize the statistical uncertainty is to compute on the same 
set of configurations 
\bea
{B_S^{\msbar} (m_b)\over B_{B_s}^{\msbar} (m_b)} = 0.95(3)^{+0.00}_{-0.02} \ , 
\eea
which, when combined with the numbers from the Tab.~\ref{tab:parameters}, gives
the wanted quantity
\bea
{\cal R}(m_b) = 
-{5\over 8} \left( {m_{B_s} \over m_b(m_b) + m_s(m_b)}\right)^2 {B_S(m_b)\over B_{B_s}(m_b)}
=- 0.93(3)^{+0.00}_{-0.01}\ .
\eea
\section{Conclusion}
\label{sec:con}
In this section, we compare our results with previous lattice studies
and draw our conclusion.
\par  The first lattice calculation of 
$B_S(\mu) $ was performed in ref.~\cite{lanl}, using the 
(unimproved) Wilson fermion action. Their $B$-parameters
were obtained  in a modified 
$\msbar$ NDR  scheme,  at the scale $\mu = 2.33$~GeV, 
at a heavy-quark mass corresponding approximately to the $D_s$ mass.
By translating their result  to
the  $\msbar$  of ref.~\cite{nierste}, one finds $B^{\msbar}_S(2.33\ \gev) =
0.81(1)$ and $\tilde B^{\msbar}_S(2.33\ \gev) =0.86(1)$. 
To compare with these numbers, we have used  our
 results at $\kappa_Q = 0.1250$,  interpolated to the strange light-quark
mass  and then evolved down to
$\mu=2.33$~GeV.  We get  $B^{\msbar}_S(2.33\ \gev) = 0.80(2)$ and 
$\tilde B^{\msbar}_S(2.33\ \gev) =1.05(2)$.
$B^{\msbar}_S$ is hence in very good agreement with the value of~\cite{lanl}, 
whereas our $\tilde B^{\msbar}_S$  
is  $20$\% larger. We note, however,  that the values  $B^{\msbar}_S(2.33\ \gev) =
0.81$  and $\tilde B^{\msbar}_S(2.33\ \gev) =0.86$ 
correspond (only evolution but no extrapolation in the heavy quark mass)  to  
$B^{\msbar}_{S} (m_b)  = 0.75$~\cite{nierste} which is 
$14\%$ smaller than our number.
\par In refs.~\cite{hiroshima2,hiroshima1}, 
NRQCD has  been used to compute $B^\prime_S(m_b)$. 
A comparison with them is interesting, because effective theories have 
different
systematic errors with respect to the approach followed in the present study.
Their latest  result  (which we convert to  $B_S$) 
reads  $B^{\msbar}_S(m_b)=0.78(2)(10)$, which is in fair agreement with ours. 
Moreover their ratio $B_S^{\msbar} (m_b)/ B_{B_s}^{\msbar} (m_b)
\sim 0.78/0.85=0.92$ is very close to ours. This implies that, in spite of the
difference in the single $B$-parameters, 
 the same physical answer for
 $\left(\Delta\Gamma_{B_s}/\Gamma_{B_s}\right)$  is obtained from 
eq.~(\ref{eq:present}).
In~\cite{hiroshima2,hiroshima1}, in order to get the width difference,
they adopted, however,  another expression
(eq.~(4) of \cite{hiroshima1}), which uses different inputs, namely
the experimental inclusive  semileptonic  branching ratio and
the theoretical determination of the decay constant $f_{B_s}=245\pm 30$~MeV. 
Their result~\cite{hiroshima2}, 
$\left(\Delta\Gamma_{B_s}/\Gamma_{B_s}\right)=(10.7 \pm 2.6\pm 1.4\pm 1.7)
\times 10^{-2}$ is about a factor of two larger than ours.
One may wonder whether the difference is due to a different evaluation
of the $1/m$ corrections, which are so important in this game.
From our calculation, we find that this contribution to 
$\left(\Delta\Gamma_{B_s}/\Gamma_{B_s}\right)$ is about
$-8.6 \times 10^{-2}$,
identical to the value used in~\cite{hiroshima2} and~\cite{hiroshima1}
(eq.~(9) of \cite{hiroshima1}). Besides the fact that they use
a formula which involves the inclusive semileptonic branching ratio,
the main difference stems, instead, from the use of a very large value
of $f_{B_s}$ (and to a lesser extent from the use of $m_b=4.8$~GeV instead
of $4.6$~GeV).   We do not find it justified to use the {\it unquenched} value
of $f_{B_s}$, combined with $B$ parameters computed in the quenched
approximation. What really matters is the combination of these
quantities in the matrix elements, and we do not know how much the $B$
parameters change in the unquenched case.  This is the reason
why we prefer  eq.~(\ref{eq:present}), which does not require
$f_{B_s}$, but  only ${\cal R}(m_b)$ (which is essentially the same
for us and in \cite{hiroshima2}) and
$\xi$,  which is known to  remain almost the same in the quenched and unquenched
case~\cite{reviews}.

On the lattice,
physical quantities relevant in heavy-quark physics can be computed
following two main routes, either by extrapolating the results
in the heavy-quark mass from a region around the charm mass
or by using some effective theory (HQET or NRQCD).
The two approaches have different systematics and in many cases
lead to results which are barely compatible.
The $B_s$ width difference is particularly lucky, in this respect,
since  ${\cal R}(m_b)$ in our calculation  and in  
ref.~\cite{hiroshima2} are in excellent agreement and  lead to the same
value of $\left(\Delta\Gamma_{B_s}/\Gamma_{B_s}\right)$ 
if eq.~(\ref{eq:present}) is used.
It is not surprising that ref.~\cite{hiroshima2} predicts  a much larger value 
for $\left(\Delta\Gamma_{B_s}/\Gamma_{B_s}\right)$, since they use
a very large value of $f_{B_s}$, which is not needed in eq.~(\ref{eq:present}).  
We find that, in order to reduce the present error, a better
determination of the $1/m$ correction, although rather hard, is very important.
Obviously, calculations with larger heavy-quark masses and without
quenching are also demanded for a better control of the remaining 
systematic errors. 

\section*{Acknowledgements}
We  warmly thank E.~Franco and O.~Schneider,
for  discussions. 
V. G. has been supported by CICYT under the Grant AEN-96-1718,
by DGESIC under the Grant PB97-1261 and by the Generalitat Valenciana
under the Grant GV98-01-80.  We acknowledge the M.U.R.S.T.
and the INFN for support.

\end{document}